\documentclass[onecolumn]{emulateapj}

\usepackage{graphicx}

\shorttitle{Grain Alignment}
\shortauthors{Cho \& Lazarian}

\begin{document}

\title{Grain Alignment by Radiation in Dark Clouds and Cores}

\author{Jungyeon Cho}
\affil{Dept. of Astronomy and Space Science, Chungnam National University, Daejeon, Korea; jcho@cnu.ac.kr}
\and
\author{A. Lazarian}
\affil{Astronomy Dept., Univ.~of Wisconsin, Madison, 
       WI53706, USA; lazarian@astro.wisc.edu}

\begin{abstract}
We study alignment of grains by radiative torques.
We found steep rise of radiative torque efficiency as grain
size increases. This allows larger grains that are known to
exist within molecular clouds to be aligned by the attenuated
and reddened interstellar radiation field. In particular,
we found that, 
even deep inside giant molecular clouds,  e.g.  at optical
depths corresponding to
$A_V \lesssim 10$, 
large grains can still be aligned by radiative torque. This means
that, contrary to earlier claims, 
far-infrared/submillimeter polarimetry  provides a reliable
tool to study magnetic fields of pre-stellar cores.
Our results show that the grain size distribution is important 
for determining the relation between the degree of polarization and intensity.

\end{abstract}

\section{Introduction}

It is widely believed that magnetic fields play a crucial role
for the dynamics of molecular clouds and for the star formation
processes (see review by Crutcher 2004 and references therein).\footnote{Existing claims of the contrary (see Padoan \& Norlund 1998) make quantitative studies of 
magnetic fields more essential.}
One of the most informative techniques of studying magnetic 
fields in molecular clouds is based on the use of starlight polarization and 
polarized emission arising from aligned dust.

Alignment of interstellar dust was not expected by theorists. 
Very soon after the 
discovery of the interstellar origin of
starlight polarization by Hall (1949) and Hiltner (1949),
it became clear that
interstellar grains get aligned with respect to magnetic field.
It did not take long time to realize that grains tend to be aligned
with their long axes perpendicular to magnetic field.
However, progress in theoretical understanding of the alignment
has been surprisingly slow in spite of the fact that great minds like
L. Spitzer and E. Purcell worked on grain alignment (see Spitzer \& Tukey 1951,
Purcell 1969, Spitzer \& Purcell 1971, Purcell 1975, 1979, 
Spitzer \& McGlyn 1979). 
The problem happened to be 
very tough and a lot of relevant physics had to be uncovered. An extended
 discussion of
different proposed mechanisms with the relevant references
can be found in a recent review by Lazarian (2003).

Originally it was widely believed that interstellar grains can be well aligned
by a paramagnetic mechanism (Davis \& Greenstein 1951). This mechanism, based
on the direct interaction of rotating grains with the interstellar 
magnetic field, required magnetic fields that are substantially
stronger than those
uncovered by other techniques.\footnote{As discussed, for instance in 
Lazarian (2003),
the very small grains are likely to be aligned by this mechanism and 
this can explain
the peculiarities of the UV part of the spectrum of the polarized 
radiation observed (see Kim \& Martin 1995).
The alignment of small grains by paramagnetic relaxation is possible
as efficiency of the Davis-Greenstein mechanism increases 
as the grain size decreases. The degree of alignment of small grains
provides a direct constrain on the intensity of magnetic field, which is
a subject that calls for more of UV polarimetry work.}
Later, a pioneering work by Purcell (1979, henceforth P79) 
showed a way how to make grain 
alignment more
efficient. Purcell noticed that grains rotating at high rates are 
less susceptible
to the randomization induced by gaseous collisions, while paramagnetic
alignment would proceed at the same rate. He introduced several 
processes that
are bound to make grains very fast "suprathermal" rotators. They are a)
variations of the accommodation
coefficient for atoms and molecules bouncing from the grain surface,
b) variations of the coefficient of electron ejection, c) variations
of the sites of H$_2$ formation over grain surface. As H$_2$ formation
over grain surfaces is a common interstellar process and every formation event
could
deposit an appreciable angular momentum with the grain, Purcell identified
c) as the major cause of grain fast rotation. He claimed that 
the catalytic sites
 ejecting H$_2$ molecules ( frequently called  ``Purcell's 
rockets'') should spin up grains very efficiently for most
of the diffuse ISM. One can easily see that within the Purcell's model,
 even a small fraction of atomic hydrogen present in molecular
clouds would also make them suprathermal (i.e. $E_{kin}\gg kT_{grain}$).   
For decades this became a standard
explanation for grain alignment puzzle, although it could not explain 
several observational
facts, e.g. why observations indicate that small grains are less aligned 
than the large ones.

New physics of grain internal motion uncovered fairly recently 
explains why  small grains are poorly aligned by Purcell's mechanism.
The inefficiency stems from grain internal wobbling. Indeed,
for sufficiently small grains it is impossible to 
assume  that they rotate perfectly
about their axis of maximal inertia. It is interesting to recall
that  the issue of grain wobbling was a part of the alignment process
discussed e.g. by Spitzer (see Jones \& Spitzer 1968). However, when
P79 identified Barnett relaxation as
an fast process of internal relaxation
that aligns  grain rotation with the axis of the maximal moment
of inertia\footnote{Such a rotation corresponds to the minimum of
grain energy for a fixed angular momentum.}, an idea
that grains {\it always} rotate about the axis
of maximal inertia got universally accepted. 
The flaw with this reasoning was found in Lazarian (1994),
where it was shown that thermal fluctuations within grain material 
induce grain wobbling, the amplitude of which depends on
the ratio of the grain rotational energy $E_{kin}$ and $kT_{grain}$.\footnote{
It is worth noting that the amplitude of wobbling does not
decrease as the efficiency of relaxation increases. The coupling
between the rotational and vibrational degrees of freedom established
by the relaxation mechanism acts back to induce wobbling well in
accordance with Fluctuation-Dissipation Theorem (see Landau \& Lifshitz 1951).
More discussion of this point is given in Lazarian \& Yan (2003).} 
The quantitative theory of the effect presented in Lazarian \& Roberge (1997)
allowed to revise the Spitzer \& McGlynn (1979) theory of crossovers (Lazarian
\& Draine 1997) as well as the theory of paramagnetic alignment of thermally 
rotating grains (Lazarian 1997a, Roberge \& Lazarian 1999).   

However, a more interesting development came about later, when
Lazarian \& Draine (1999a, henceforth LD99a) realised that grains not only wobble but
occasionally flip. LD99a found that small grains flip more frequently
that large ones.
As the result
regular torques, e.g. torques due to ejection of H$_2$ molecules, 
get averaged out over flipping grains and they get "thermally trapped",
i.e. rotate at thermal velocities in spite of the presence of Purcell's 
torques. Taking into account that the paramagnetic alignment of thermally
rotating grains is rather inefficient  (see Roberge \& Lazarian 1999)
it is possible became possible to explain why small grain may
be poorly aligned.

A new twist to the theory of grain alignment came about when
Lazarian \& Draine (1999b, henceforth LD99b) 
found that species with nuclear moments
within the grain, e.g. $^1$H, $^{13}$C, $^{27}$Al, $^{31}$P, 
$^{29}$Si, $^{55}$Mn... , bring about a new type of
internal relaxation which was termed "nuclear relaxation".
This type of relaxation for grains larger than $10^{-5}$~cm
happened to be $\sim 10^6$ times more
efficient that the Barnett relaxation introduced in Purcell (1979).
As the result, LD99b claimed that for diffuse interstellar gas 
nuclear relaxation
thermally traps grains of the sizes from $10^{-5}--$~cm 
making the Purcell mechanism inefficient. 

A group of alternative mechanisms of alignment that rely on  the relative 
gas-grain
motion have their particular niches. The first mechanical
alignment mechanism was pioneered by Gold (1951). Later work included 
driving grains
by ambipolar diffusion (Roberge \& Hanany 1990, 
Roberge, Hanany \& Messinger 1995) and Alfven waves 
(Lazarian 1994, Lazarian 1997b, 
Lazarian \& Yan 2002, 
Yan \& Lazarian 2003). Although new efficient processes of
mechanical alignment have been proposed
(Lazarian 1995, Lazarian \& Efroimsky 1996, Lazarian, Efroimsky \& Ozik 
1996), this did not make
mechanical alignment universally applicable. 
 
All this provided the background that made radiative torques mechanism 
most promising for explaining grain
alignment over vast expanses of the interstellar space. Introduced first by
Dolginov (1972) and Dolginov \& Mytrophanov (1976) the radiative torques were
mostly forgotten till the pioneering work by Draine \& Weingartner 
(1996; hereafter DW96), where their efficiency was demonstrated
using numerical simulations.
The radiative torques make use of interaction of radiation with a grain 
to spin it up. Indeed, in general, one would expect that the
cross sections of the interaction of an irregular grain with left and
right circular polarized photons are different. As the non-polarized light
can be presented as a superposition of the equal fluxes of photons with
opposite circular polarization, the interaction of such a light with
the irregular grain would result in grain spin up. Unlike Purcell's
torques, that are fixed in grain frame, the radiative torques are 
expected to be less affected by grain flipping.

The predictions of radiative torque mechanism are roughly 
consistent
with the molecular cloud extinction and emission polarimetry 
(Lazarian, Goodman \& Myers 1997)
and the polarization spectrum measured (see Hildebrand et al. 2000). 
They have been
demonstrated to be efficient in a laboratory setup (Abbas et al. 2004). 
Evidence in favor of radiative
torque alignment was found for the Whittet et al. (2001) 
data obtained at the interface of the
dense and diffuse gas at the Taurus cloud (see Lazarian 2003).

In view of this success the radiative torque mechanism is the primary 
mechanism that we are going to study
in relation to grain alignment deep inside molecular clouds.
While a possible failue of radiative torques there does not exclude
that grains are aligned deep within molecular clouds, their
success would definitely make polarimetric studies of molecular
clouds much more trustworthy and informative.
Whether grains are aligned there is necessary to understand to know
whether aligned grains trace only surface magnetic fields 
or magnetic fields deeply embedded into molecular clouds.
The topology of magnetic field inside molecular clouds
is essential for understanding for star formation.

It has been shown that optical and near infrared polarimetry 
provide magnetic fields 
only to $A_v$ of 2 or less (Goodman et al. 1995, Acre et al. 1998). 
Is it the same for far infrared
polarimetry? This is the question that we address in this paper. Earlier 
answers
(see Lazarian et al. 1997) appeal to stars embedded in the cloud. Indeed,
such stars can induce alignment through their radiation. Here we consider an
extreme case of a cloud without any embedded stars. This situation is also
motivated observationally, as some recent observations indicate that there
are aligned grains deep in molecular clouds without high mass stars 
 (Ward-Thompson et al. 2000).
 
In what follows we discuss grain alignment by radiative torque in
molecular clouds. In \S2, we calculate efficiency of radiative torque 
in a molecular cloud and minimum aligned grain sizes as a function
of visual extinction in the cloud.
In \S3, we calculate polarized far-infrared/submillimeter
emission from a pre-stellar core and
discuss the relation between the degree of polarization and
intensity. We give discussion in \S4 and conclusion in \S5.

\section{Radiative Torques}

As we mentioned above, a
 flow of photons illuminating a grain can be presented as a superposition
of left- and right- handed photons, while an irregular grain has different
cross section of interaction with photons of different handedness. As
the result of differential extinction, i.e. absorption and scattering,
 the grain feels a regular torque. Note, that the key word here is
regular. Random torques produced photons emitted and absorbed
 by a grain  
were discussed in terms of grain alignment by Harwit (1970). They are
rather inefficient, however (Purcell \& Spitzer 1971) and are more
important in terms of damping of grain rotation (see Draine \& Lazarian 1998).

Although the physics of grain spin up by radiative torques was  for the
most part properly understood
 by Dolginov \& Mytrophanov (1976),
only calculations in DW96 provided a
quantitative insight into the process. 
These calculations 
obtained for test grains
using Discrete Dipole Approximation code (Draine \& Flatau 1994) showed that
both anisotropic and isotropic radiation flows can efficiently spin-up
grains. While Dolginov \& Mytrophanov (1976) did understand that grains
will not be only spun up, but also aligned by anisotropic radiation,
they could not get correctly what would be such an alignment.
Numerical simulations In Draine \& Weingartner (1997)
reveal complex dynamics of grains and revealed
that in most cases the grains {\it tend} to get aligned
with long axes perpendicular to magnetic field, even if
paramagnetic relaxation is absent. 
 Although the nature of this alignment
in the absence of analytical calculations still remains unclear and some
features of grain internal dynamics (that were discovered later!) are
missing in the model studied 
(see an attempt in this direction in Weingartner \& 
Draine 2003), it is very plausible that radiative torques can provide
the alignment that corresponds to observations. Appealing
to polarimetric data available one can claim that observations
do not give us any indications that anisotropic
radiation provides alignment that either has wrong sign or depends
on the angle between magnetic field and anisotropy direction.
This would be the case, however, if the dynamics of interstellar
grains were different from the assumed one. For the rest of the paper
we assume that the radiative torques do align grains with long axes
perpendicular to magnetic field and will concentrate therefore only
on the magnitude of radiative torques.

While calculations in DW96 were
limited by the interstellar grains, we study radiative alignment of
grains of larger sizes. Such grains are known to be present in molecular
clouds. In addition, unlike DW96, here we are interested in the alignment of 
grains by attenuated
and reddened interstellar light that enters into a cloud from outside.

\subsection{Method}
We use the DDSCAT software package 
(astro-ph/0309069; DW96)
to calculate radiative torque on grain particles.
We use the same grain shape as in DW96, 
which is an asymmetric assembly of 13 identical cubes.
The grain is subject to radiative torque because of its irregular shape.
We use the refractive index of astronomical silicate (Draine \& Lee 1984; 
Draine 1985;
Loar \& Draine 1993; see also Weingartner \& Draine 2001).

In our calculations, the incoming radiation is parallel to the
principal axis ${\bf a}_1$ of the grain. 
Therefore the target orientation angle $\Theta$,
the angle between the incident
radiation and the grains primary axis $\hat{\bf a}_1$ (see DW96), 
is zero.
Therefore in our calculations
the radiative torque, ${\bf \Gamma}_{rad}$, is parallel to $\hat{a}_1$ and
$|{\bf \Gamma}_{rad}|=|{\bf \Gamma}_{rad}\cdot \hat{\bf a}_1|$.

\begin{figure*}[h!t]
\includegraphics[width=.48\textwidth]{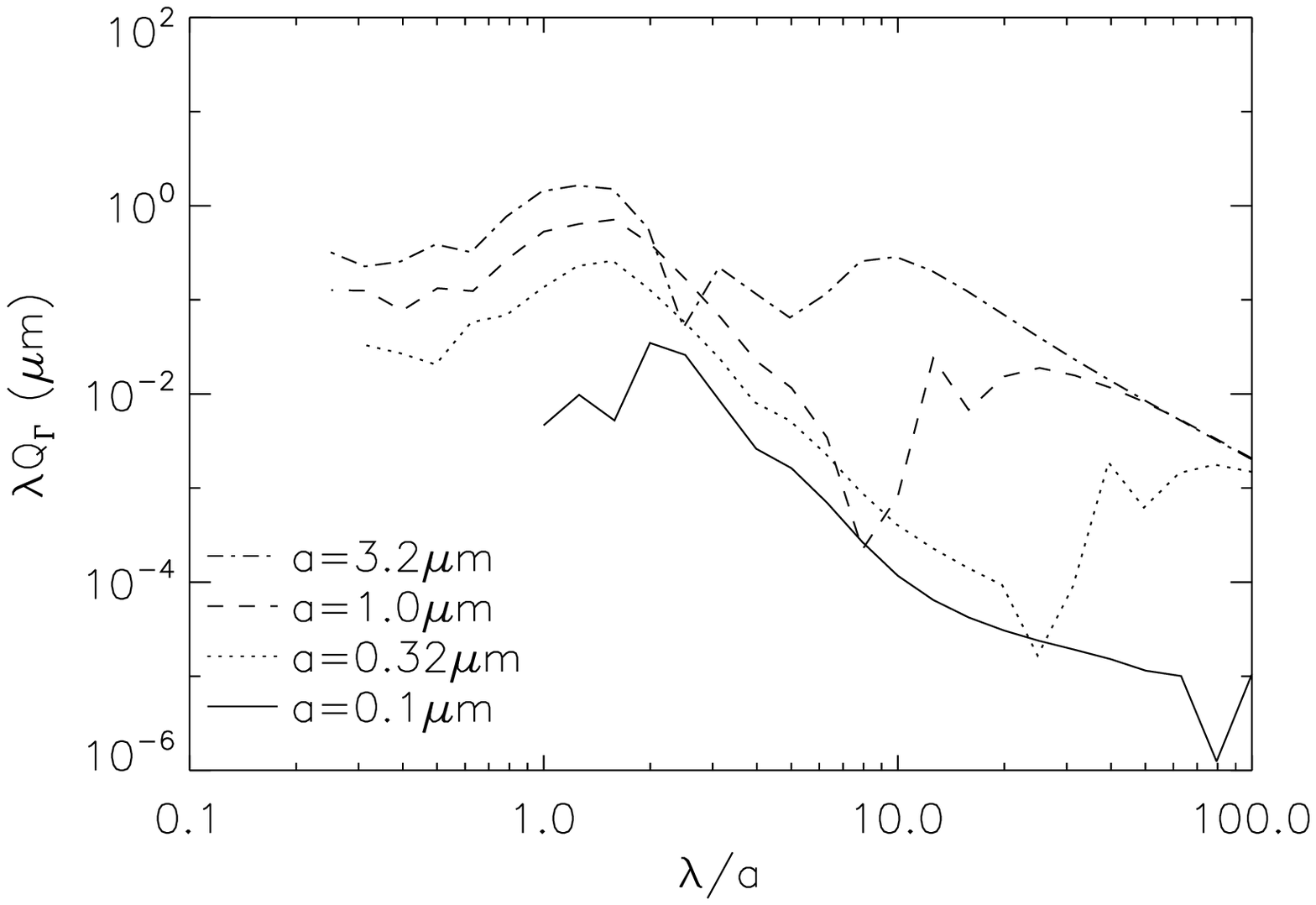}
\includegraphics[width=.48\textwidth]{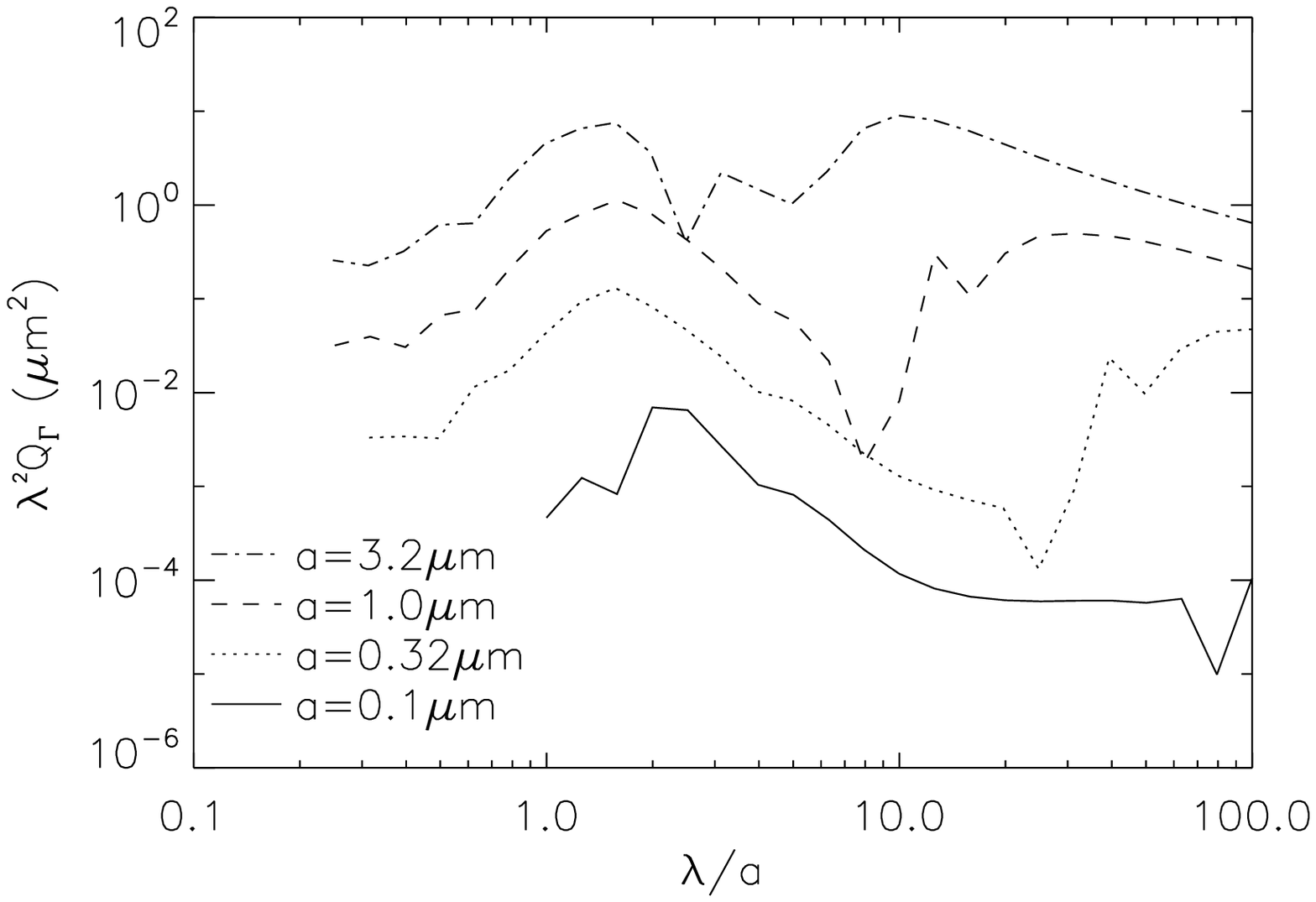}
\caption{
 Dependence of $\lambda |Q_{\Gamma}\cdot \hat{\bf a}_1|$ (left panel) 
 and $\lambda^2 |Q_{\Gamma}\cdot \hat{\bf a}_1|$ (right panel) on 
 $\lambda/a$, where
 $\lambda$ is the wave length and $a$ is the grain size.
 These quantities are useful for estimating 
 which part of the electromagnetic
 spectrum contributes most to the radiative torque.
\label{fig:1}
}
\end{figure*}

For a given wavelength and a grain size, the DDSCAT package returns 
the torque efficiency vector ${\bf Q}_{\Gamma}$:
\begin{equation}
 {\bf Q}_{\Gamma} \equiv \frac{ {\bf \Gamma}_{rad} }{ \pi a_{eff}^2 u_{rad}
  \lambda/2\pi },
\end{equation}
where ${\bf \Gamma}_{rad}$ is the radiative torque, 
$a_{eff}\equiv (3V/4\pi)^{1/3}$ the effective target radius, 
$V$ the volume of the target, 
$u_{rad}$ the energy density of the incident radiation,
and $\lambda$ the wavelength.
When we consider a radiation field with the mean intensity $J_{\lambda}$,
the radiation torque becomes
\begin{equation}
  {\bf \Gamma}_{rad}=\pi a_{eff}^2   
 \int~\mbox{d}\lambda (4\pi J_{\lambda}/c)\frac{\lambda}{2\pi} {\bf Q}_{\Gamma},
  \label{eq:tq_lin}
\end{equation} 
where we used $u_{rad}= 4 \pi J_{\lambda}/c$.
When we perform the summation over $\lambda$-axis in natural logarithmic scale,
the summation becomes
\begin{equation}
  {\bf \Gamma}_{rad}=2.303 \Delta(\log_{10} \lambda)( a_{eff}^2 /2c )  
 \sum_i (4\pi J_{\lambda,i}) \lambda_i^2 {\bf Q}_{\Gamma,i}, \label{eq_Gamma}
\end{equation}
where we used $\Delta(\log_{10}\lambda)=2.303~d\lambda / \lambda$.
Figure \ref{fig:1} shows that the value of 
$\lambda_i |{\bf Q}_{\Gamma,i}|$ 
$(=\lambda_i |{\bf Q}_{\Gamma,i}\cdot \hat{\bf a}_1|)$ and
$\lambda_i^2 |{\bf Q}_{\Gamma,i}|$ 
$(=\lambda_i^2 |{\bf Q}_{\Gamma,i}\cdot \hat{\bf a}_1|)$
as a function of $\lambda/a$ for large grains.
The quantity $\lambda_i |{\bf Q}_{\Gamma,i}|$ 
is useful for
integration in equation (\ref{eq:tq_lin}) and
$\lambda_i^2 |{\bf Q}_{\Gamma,i}|$ for
integration in equation (\ref{eq_Gamma}).

Mathis, Mezger, \& Panagia (1983) showed that the average interstellar
radiation field (ISRF) in the solar neighborhood consists of
a small UV component plus three blackbody components with
$T$ = 3000, 4000, and 7500 K. The blackbody components are given by
\begin{equation}
 4\pi J_{\lambda} \lambda =\sum_j W(T_j) 4\pi \lambda \frac{2 h c^2}{\lambda^5}
\frac{1}{ exp(hc/\lambda kT_j)-1 },
\end{equation} 
where $W(T=3000)=4\times 10^{-13}, W(T=4000)=1.65\times 10^{-13}, 
W(T=7500)=1\times 10^{-14}$, $k=1.38\times 10^{-16}$, and $h=6.63\times 10^{-27}$ in cgs units.
See Figure 1 of Mathis et al.~(1983).
Radiation field inside a giant molecular cloud located at $r_G=5kpc$ is 
also given in Figure 4 of Mathis et al.~(1983).  
They considered a spherical Giant molecular cloud that has
an isotropic radiation (i.e.~ISRF) incident upon the surface of the
cloud. They produced the mean radiation intensity $J_{\lambda}$ as a function
of the visual extinction $A_V$ measured from the surface
of an opaque cloud.
We calculate radiative torque inside a giant molecular cloud located at
$r_G=5kpc$ using the radiation field given in Mathis et al.~(1983).

Once we know ${\bf Q}_{\Gamma,i}$ (from DDSCAT) and $J_{\lambda,i}$ (from
Mathis et al. 1983), we can obtain the torque from 
equation (\ref{eq_Gamma}).
The gas drag damps grain angular rotation. The gas drag torque is given
by
\begin{equation}
      |{\bf \Gamma}_{drag,gas}\cdot \hat{\bf a}_1|=(2/3)\delta n_H
     (1.2)(8\pi m_H kT)^{1/2} a_{eff}^4 \omega,    
\end{equation}
where $n_H$ is the hydrogen number density, 
$m_H$ is the mass of a hydrogen atom, $\delta \approx 2$, 
and $\omega$ is the angular frequency (see DW96).
By equating the radiative torque $|{\bf \Gamma}_{rad}|$ 
$(=|{\bf \Gamma}_{rad}\cdot \hat{\bf a}_1|)$
in equation (\ref{eq_Gamma}) and
the gas drag torque $|{\bf \Gamma}_{drag,gas}\cdot \hat{\bf a}_1|$ above,
we can
obtain the angular velocity of grain rotation around $\hat{\bf a}_1$:
\begin{equation}
  \omega_{rad} = \frac{ |{\bf \Gamma}_{rad}| }
        { (2/3)\delta n_H (1.2)(8\pi m_H kT)^{1/2} a_{eff}^4 }
        \left( \frac{1}{ 1+ \tau_{drag,gas}/\tau_{drag,em} } \right),
  \label{eq:omegarad}
\end{equation}
where values and definitions of the gas drag time and thermal emission
drag time, $\tau_{drag,gas}$ and 
$\tau_{drag,em}$ respectively, are given in DW96\footnote{Additional
processes, e.g. plasma drag, were discussed in Draine \& Lazarian (1998). 
These processes are essential for small grains, but less important for
large grains that we primary deal with here.}.

The thermal rotation rate $\omega_T$ is the rate at which
the rotational kinetic energy of a grain is equal to $kT/2$:
\begin{equation}
  \omega_T= \frac{ 15 kT }{ 8 \pi \alpha_1 \rho a_{eff}^5 }.
  \label{eq:omegat}
\end{equation}
When a grain rotates much faster than $\omega_T$,  the randomization
of a grain by gaseous  collisions is reduced.
Therefore, if a grain rotates superthermally, we expect 
the grain rotation axis $\hat{\bf a}_1$ is aligned 
with magnetic field.
{}From equations (\ref{eq:omegarad}) and (\ref{eq:omegat}), we have
\begin{equation}
  \left( \frac{\omega_{rad}}{\omega_T} \right)^2 =
  \left( \frac{ |{\bf \Gamma}_{rad}| }
        { (2/3)\delta n_H (1.2)(8\pi m_H kT)^{1/2} a_{eff}^4 } \right)^2
  \left( \frac{ 8 \pi \alpha_1 \rho a_{eff}^5 }
              { 15 kT } \right)^2
  \left( \frac{1}{ 1+ \tau_{drag,gas}/\tau_{drag,em} } \right)^2,
\end{equation}
or,
\begin{equation}
  \left( \frac{\omega_{rad}}{\omega_T} \right)^2 =
  \frac{ 5\alpha_1 }{ 192 \delta_2 }
  \left( \frac{ 1 }{ n_H kT } \right)^2
  \left( \frac{\rho a_{eff}}{m_H} \right)
 \left( \gamma \int d\lambda {\bf Q}_{\Gamma} \lambda (4\pi J_{\lambda}/c) \right)^2
  \left( \frac{1}{ 1+ \tau_{drag,gas}/\tau_{drag,em} } \right)^2,
\end{equation}
where $\gamma$ is the anisotropy factor of the radiation field.
We use $\gamma=0.1$ for diffuse cloud and $\gamma=0.7$ for the GMCs
in accordance with DW96.
When the ratio is larger than 1, radiative torque is an efficient 
mechanism for grain alignment.

\subsection{Results}
\begin{figure*}[h!t]
\includegraphics[width=.48\textwidth]{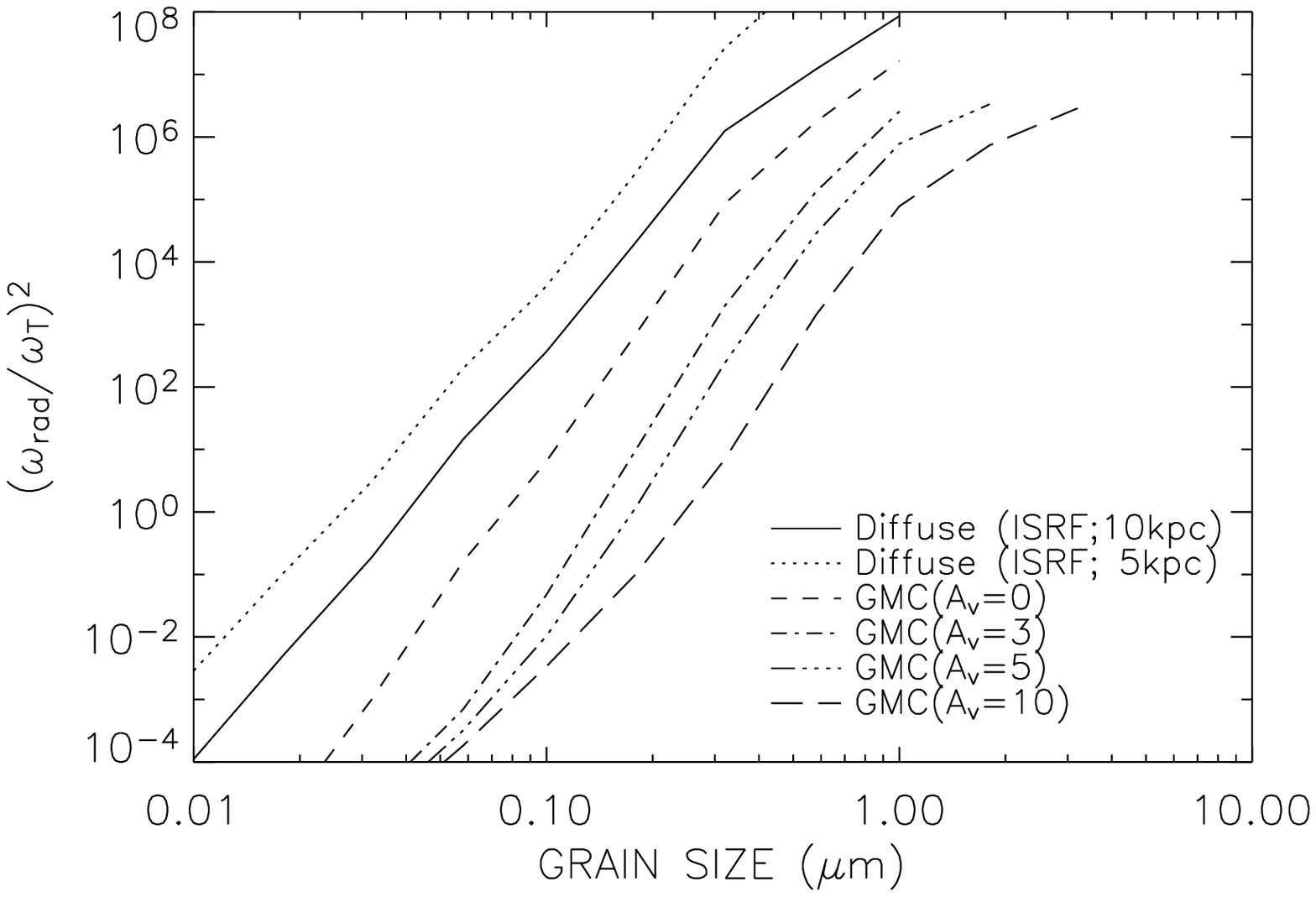}
\includegraphics[width=.46\textwidth]{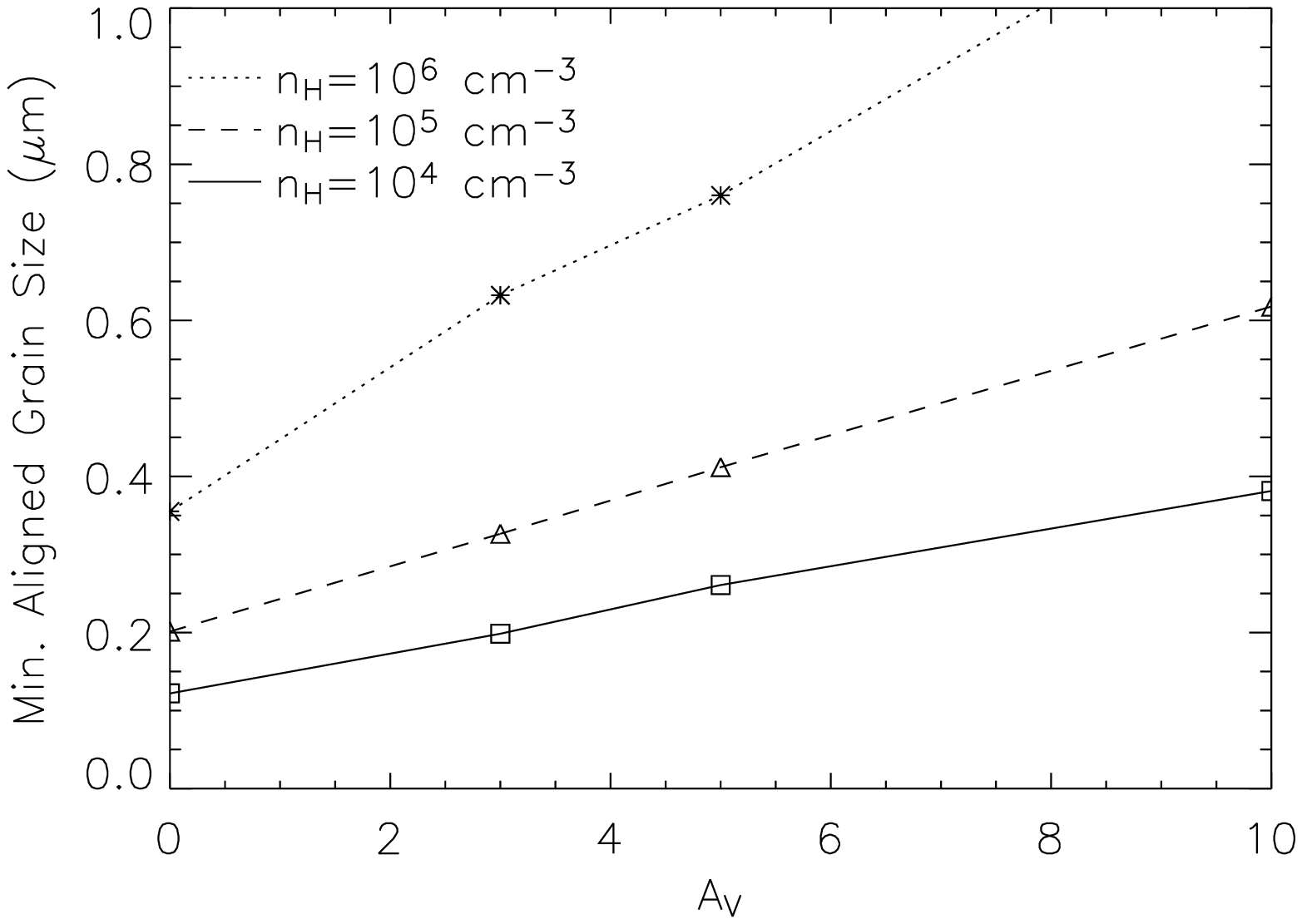}
\caption{
    Efficiency of radiative torque. 
    When $\omega_{rad}/\omega_T > 1$, radiative torque can
    rotate grains superthermally, which results in
    grain alignment.
    Different curves represent radiative
    torque by different radiation fields. The visual extinction $A_V$
    is for a giant molecular cloud located at 5kpc from
    the Galactic center. We assume $n_H=10^4 cm^{-3}$ and $T=20K$ for
    the GMC (see Table 6 in DW96 for
    other parameters). For diffuse ISM, we use $n_H=30 cm^{-1}$ and
    $T=100K$ (see Table 5 in DW96).
 \label{fig:2}
}
\caption{
  Minimum aligned grain size vs. visual extinction $A_V$.
  We use $T=20K$ parameters given in DW96. 
  However, note that we consider 3 different densities. 
\label{fig:3}
}
\end{figure*}

We show the results for $(\omega_{rad}/\omega_T)^2$ in Figure \ref{fig:2}.
The solid line is for the interstellar radiation field (ISRF) in the solar
vicinity (see Mathis et al. (1983) for details about the radiation field).
DW96 used this radiation field and obtained
the $\omega_{rad}/\omega_T$ ratio for three grain sizes ($a_{eff}=0.02,
0.05$, and $0.2\mu$m.).
In their calculation,
they included both isotropic and anisotropic components of
radiation field. 
Our calculations are slightly different.
Indeed, we consider only anisotropic radiation component
the effect of which on alignment is substantially 
stronger for the grain that we use 
than that of the isotropic component (see Table 4 in DW96).\footnote{ We
believe that this is generally true for an ensemble of grains of
arbitrary shapes, but more studies are necessary to prove this point.
   } The calculations of radiation anisotropy in a turbulent
molecular cloud made for us by Tom Bethel show that we do not
overestimate $\gamma$s. On the contrary, these calculations
testify that in this paper, if anything, we
underestimate the actual values of radiative
torques. 

Another simplification is that we consider only anisotropy
of radiation only along magnetic field. This is justifiable for
obtaining a crude estimate, which is the actual goal of our paper.
In addition, unlike DW96, 
we use the UV smoothed refractive index of silicon (see Weingartner \&
Draine 2001).
Nevertheless, our result (solid line) agrees with that of 
DW96 within a factor of $\sim$2.


In Figure \ref{fig:3}, we show aligned grain size as a function of
visual extinction $A_V$.
We used Figure \ref{fig:2} and 
assumed that grains with $\omega_{rad}/\omega_T > 5$ are aligned.
For a cloud with $n=10^{4}$, $0.2 \mu$m grains are aligned even at
$A_v \sim 4$.  
However, for a cloud with $n=10^{5}$, $0.2 \mu$m grains are hardly 
aligned.
Grains of $\sim 1\mu$m are aligned even at $A_V\sim 10$
if the density does not exceed several times $10^5 cm^{-3}$.

In their classical paper, Mathis, Rumpl, \& Nordsieck (1977; hereafter MRN),
constructed a model for size distribution of dust grains
in the diffuse interstellar medium (ISM). 
This MRN distribution has a sharp upper cutoff at $a_{max}=0.25 \mu$m. 
The MRN model provides a good fit to interstellar
extinction and therefore widely used for modeling the diffuse ISM.
It is expected that at larger optical depths the   upper cutoff
occurs at larger values. 
For example, Kim, Martin, \& Hendry (1994) used the maximum entropy
method and obtained a smooth decrease of size distribution
starting at $0.2 \mu$m. 
Weingartner \& Draine (2001) also obtained an extended distribution
beyond the MRN upper cutoff. Physically, coagulation of grains
happens in denser parts of the interstellar gas (see discussion
in Yan \& Lazarian 2003, Yan, Lazarian \& Draine 2004).
Therefore it is resonable to  assume that grains larger than
the the usual MRN cutoff are present .
Since coagulation is more frequent in dense clouds than the diffuse ISM,
we expect to see a substantial amount of grains larger than $0.25 \mu$m
in dense clouds 
(see, for example, Clayton \& Mathis 1988; Vrba, Coyne, \& Tapia 1993).
If grains of $\sim 1\mu$m are abundant in dark clouds, they can 
emit polarized infrared radiation even deep inside the cloud.

\subsection{Polarization: Rayleigh  reduction factor}

\begin{figure*}[h!t]
\begin{center}
\includegraphics[width=.48\textwidth]{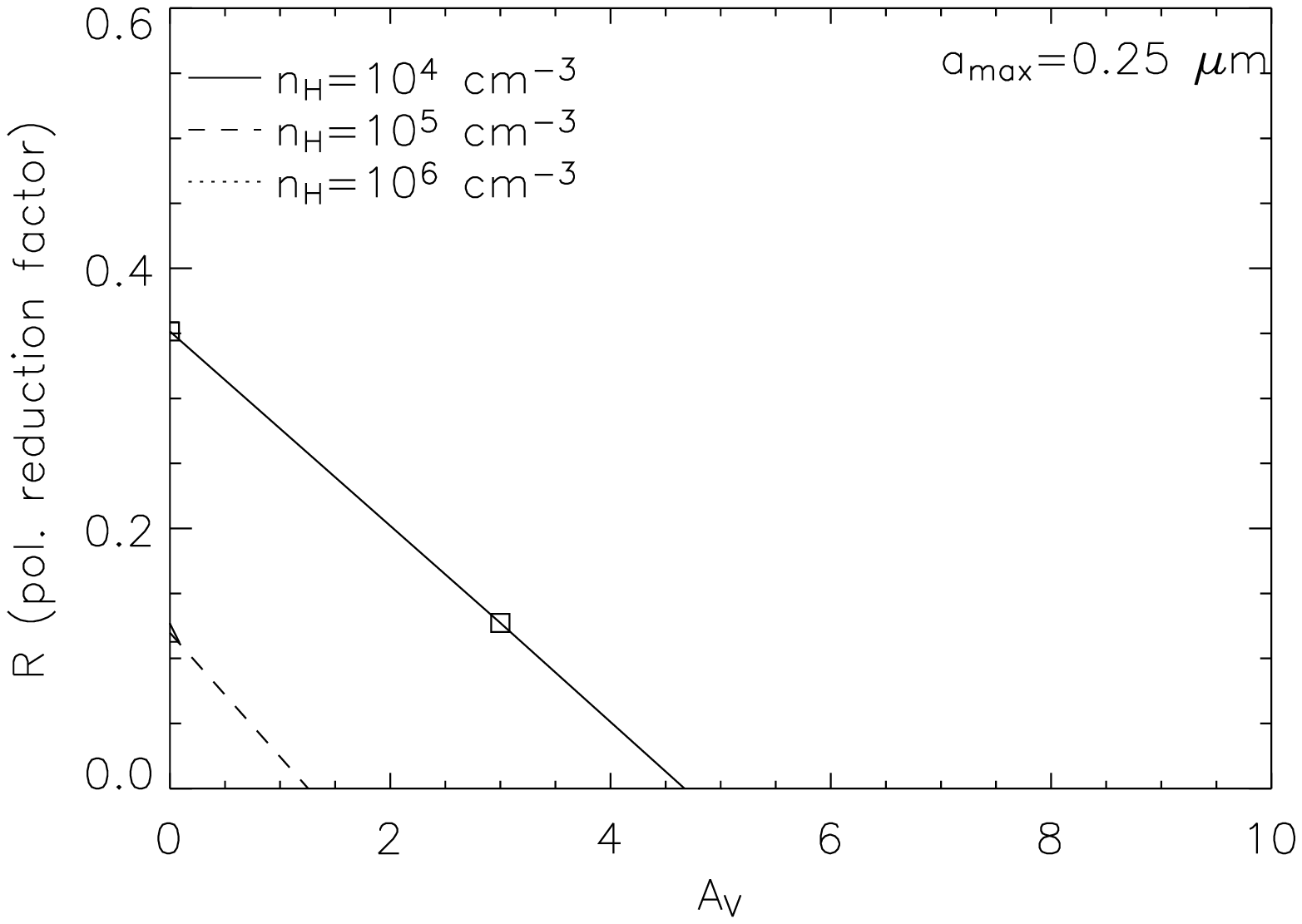}
\includegraphics[width=.48\textwidth]{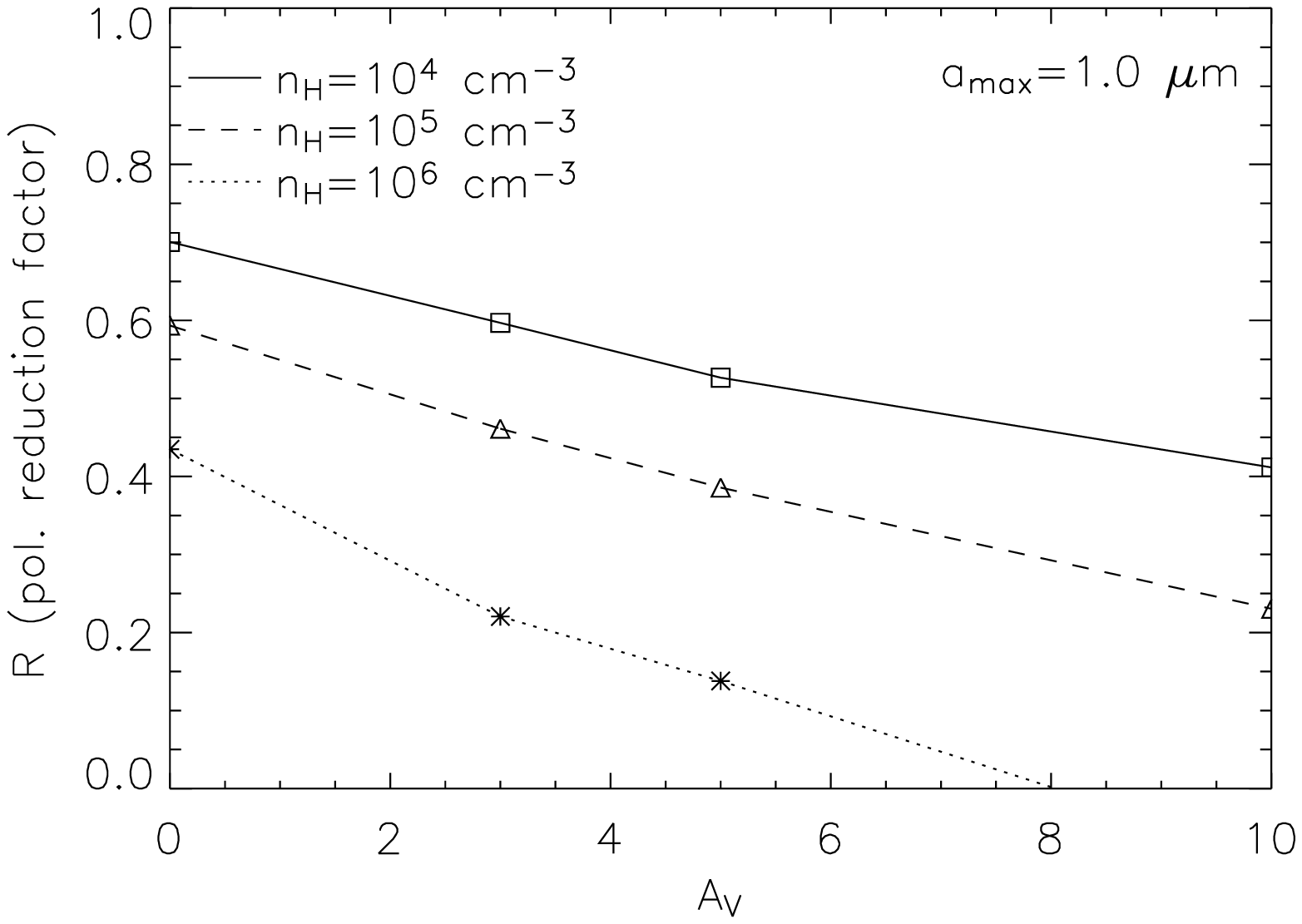} \\
 (a)  \hspace{70mm}  (b)
\end{center}
\caption{
   Rayleigh polarization reduction factor 
  (see equation (\ref{eq:R}) for the definition in our case).
  (a) The original MRN distribution with $a_{max}=0.25 \mu$m.
  (b) An extended MRN distribution with $a_{max}=1.0 \mu$m.
 \label{fig:4}
}
\end{figure*}

In Figure \ref{fig:4}, we plot the Rayleigh polarization reduction factor $R$
(p. 328 in Greenberg, 1963; see also Lee \& Draine 1985), which
is a measure of imperfect alignment of the grain axes
with respect to magnetic field. 
The conventional definition of
the factor is $R=1.5(\langle \cos^2 \beta \rangle - 1/3)$,
where $\beta$ is the angle between the grain angular momentum vector
and magnetic field.
The degree of polarization is reduced when
some grains are not perfectly aligned in respect to  magnetic field.
In our case this happens for an ensemble of grains some of which,
namely, small ones, are not aligned, while the other, namely, the large
ones, are perfectly alinged. As the polarization for the range
of far infrared $\lambda$  and grain sizes $a$ does not depend
on those parameters, we can calculate
the reduction factor for the entire distribution of grains as follows:
\begin{equation}
  R=\frac{ \int_{a_{aligned}}^{a_{max}} C_{ran} n(a) da }
         { \int_{a_{min}}^{a_{max}} C_{ran} n(a) da },
  \label{eq:R}
\end{equation}
where $C_{ran}$ is the cross-section, $n(a)$ the grain number density,
$a$ the grain size, $a_{min}$ the minimum size of grains,
$a_{max}$ the maximum size, and
$a_{aligned}$ the minimum aligned size, which is given in
Figure \ref{fig:3}.
We assume MRN-type power-law grain size distributions:
\begin{equation}
  n(a) \propto a^{-3.5}
\end{equation}
for $a=0.005\mu$m to $a=a_{max}$ $\mu$m.
We consider two values for $a_{max}$: the original MRN cutoff at $a_{max}=0.25\mu$m and a larger cutoff at $a_{max}=1\mu$m
for the calculation of $R$.
We show the results in Figure \ref{fig:4}(a) and \ref{fig:4}(b), respectively.
For the original MRN distribution (Figure \ref{fig:4}(a)), 
R is smaller than
$\sim 0.4$ for $n_H > 10^4 cm^{-3}$.
However, for the larger upper cutoff (Figure \ref{fig:4}(b)),
R is about $\sim 0.2$ inside clouds at $A_V=10$
when $n_H \sim 10^5 cm^{-3}$.

We present the  results for an opaque giant molecular cloud located at 5kpc
from the Galactic center.
As we explained earlier, we used radiation field given in 
Mathis et al. (1983). They calculated the radiation field
assuming that
the visual extinction $A_V$ at the center measured from the surface is 200.
However, as long as the central visual extinction is larger than
$\sim 15$, the radiation field
may not be sensitive to the choice of 
the central $A_V$ (see Flannery, Roberge, \& Rybicki, 1980).
Therefore, our qualitative results obtained here are applicable to
various astrophysical objects - from dense prestellar cores to giant
molecular clouds.

\section{Polarized emission from a dark core}
In the previous section, we showed that larger grains can rotate
super-thermally even at $A_V=10$ in giant molecular clouds.
In this section, we apply the result to dense pre-stellar cores.
As we noted at the end of the last section, 
we obtained the results in the previous section using
the radiation field suitable for giant molecular clouds.
Therefore, it is questionable whether or not we can directly apply
the results in the previous section to prestellar cores.
However, judging from Flannery et al. (1980) calculation,
we expect that the direct application is ill-justified only near
the very center of the cores.

\subsection{Method}

In this section 
we calculate polarized emission from a dark pre-stellar core.
We assume a simple spherically symmetric density distribution and
a constant temperature ($T\sim 20K$).
We take density profile of logatropic sphere (Lizano \& Shu 1989; 
McLaughlin \& Pudritz 1996), 
which has a finite central
density and a $\rho \propto r^{-1}$ envelope.
The logatropic sphere is supported by turbulent pressure (and
isothermal gas pressure in its original form).
The turbulent pressure represents nonthermal velocity dispersion
observed in clouds.
The density profile of a logatropic sphere is different from
the critically stable isothermal Bonner-Ebert sphere 
 (Bonner 1956; Ebert 1955), 
which has a $\rho \propto r^{-2}$ envelope.
Although some observations (e.g. Alves et al. 1998; Lada, Alves, \& Lada
1999; Johnstone \& Bally 1999) support the $\rho \propto r^{-2}$ profile,
other observations (e.g. van der Tak et al. 2000; 
Colome, di Francesco, \& Harvey 1996; Henning et al. 1998)
support the other profile.
For simplicity, we use
\begin{equation}
  \rho(r) \propto \left \{ \begin{array}{ll}
                            \mbox{constant} & \mbox{if $r<r_0/4.7$} \\
                            r^{-1}          & \mbox{otherwise,}
                            \end{array}
                  \right.
\end{equation}
where $r_0$ is a parameter in our calculations (see McLaughlin \& Pudritz
1996 for its physical meaning) and we set the central
number density $n_{H,c}$ to $3\times 10^{5} cm^{-3}$.
This distribution truncates at $r\sim 24r_0$.
We take a magnetic field from our earlier direct 3-dimensional 
numerical simulation  (see Cho \& Lazarian 2003).
Numerical resolution is $216^3$ and
the average Mach number is $\sim 7$.
The magnetic field has both uniform and random components.
The strength of the mean field is about 2 times stronger than
the fluctuating magnetic field.
We assume that the uniform field is perpendicular to the line of
sight of the observer.

We assume that the visual extinction $A_v$ at the center measured from
the surface
is $\sim 10$.
This means that the total column density through the center is
about $N_H \sim 3.7 \times 10^{22} cm^{-2}$.
The size of the cloud corresponds to $\sim 0.02pc$. 
This cloud is similar to, for example, L183 (see Crutcher et al. 2004).

We assume an MRN type
grain size distribution,
  $n(a) \propto a^{-3.5}$,
from $a=0.005\mu$m to $a=a_{max}$.
Unlike the original MRN distribution, where $a_{max}=0.25\mu$m, we use
$a_{max}$ of up to 2$\mu$m.
We assume that grains are oblate spheroid 
with the axial ratio of $\sim 1.2$, which is smaller than the value
used by Padoan et al. (2001).

We follow the method somewhat modified from the 
one in Fiege \& Pudritz (2000)
to compute the polarization maps.
Here we briefly describe the procedures.
It is natural to assume (see Fiege \& Pudritz 2000)
that one can ignore the effects of absorption
and scattering when  dealing with submillimeter wavelengths.
Therefore  the polarization submm range is due to pure emission.
The Stokes parameters are given by
\begin{eqnarray}
  Q \propto C_{pol}Rq, \\
  U \propto C_{pol}Ru, \\
  I \propto C_{ran} \left[ \int \rho ds - \frac{ C_{pol}R }{ C_{ran} }
          \int \rho \left(\frac{\cos^2\gamma}{2}-\frac{1}{3} \right) ds \right],
\end{eqnarray}
where
\begin{eqnarray}
  C_{pol}=C_{\perp}-C_{\|}, \\
  C_{ran}=(2C_{\perp}+C_{\|})/3, \\
  q=\int \rho \cos^2\psi \cos^2 \gamma ds, \\
  u=\int \rho \sin^2 \psi \cos^2 \gamma ds, 
\end{eqnarray}
R is the polarization reduction factor,
$\psi$ the angle between the projection of the
local ${\bf B}$ on the plane
of the sky and north, and $\gamma$ is the angle between 
the local ${\bf B}$ and the plane of the sky.
As explained earlier, the factor $R$ is the n 
reduction factor
due to imperfect grain alignment.
Note that $a_{aligned}$ is a function of both $n_H$ and $A_V$.
Figure \ref{fig:3} shows how $a_{aligned}$ is related to $n_H$ and $A_V$.
Based on the figure we assume that 
\begin{equation}
a_{aligned}=\left[ \log_{10}( n_H ) \right]^3(A_V+5)/2800 \mbox{~$\mu$m}.
\label{eq:ali}
\end{equation}
The error of this fitting formula is around $\sim 10\%$.
Note that this fitting formula does not have any physical background.
{}From $Q, U$, and $I$, 
we can obtain polarization angle $\chi$ and the degree
of polarization as follows:
\begin{eqnarray}
 \tan 2\chi = u/q, \\
 P=\frac{ \sqrt{ Q^2 + U^2} }{ I }.
\end{eqnarray}

\begin{figure*}[h!t]
\begin{center}
\includegraphics[width=.48\textwidth]{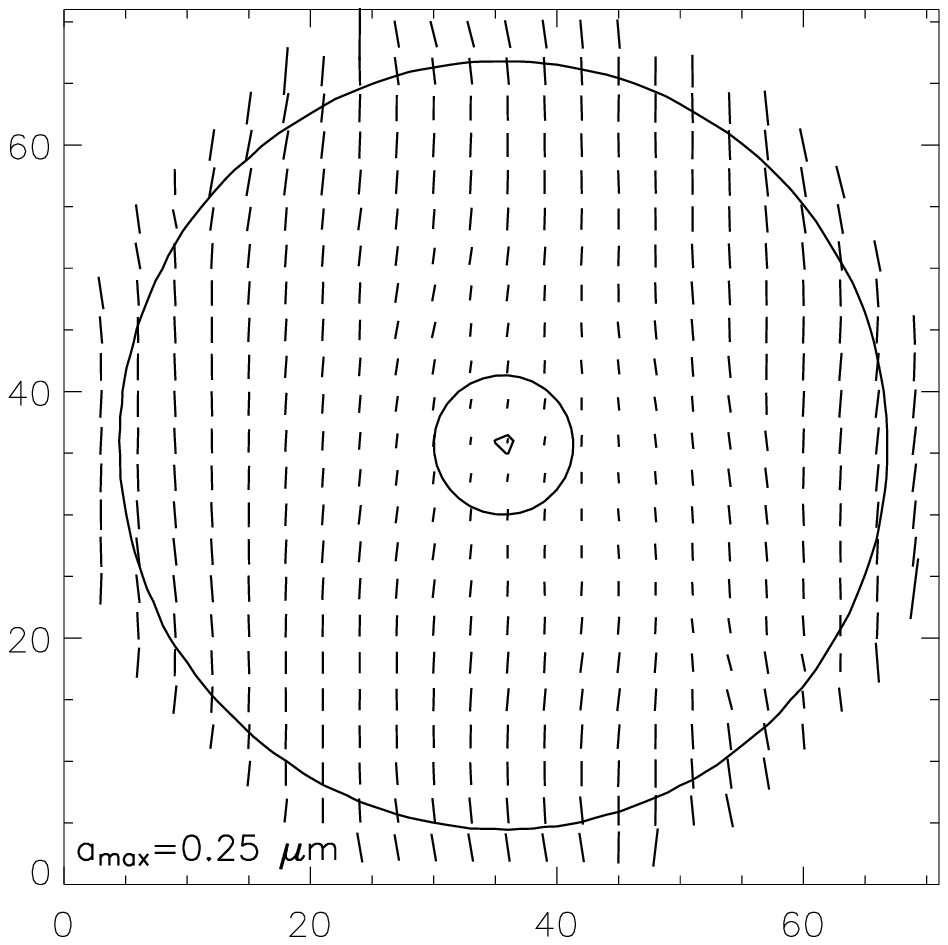}
\includegraphics[width=.48\textwidth]{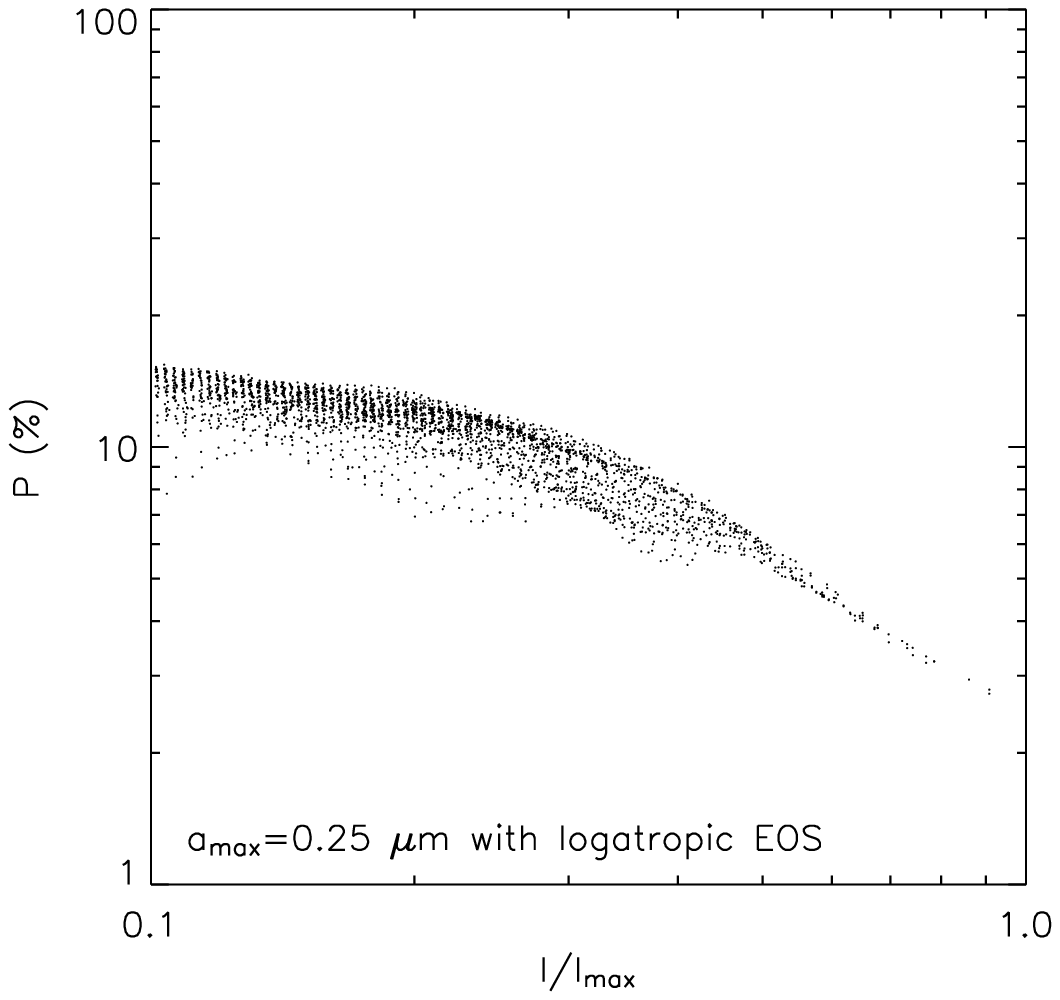} \\
 (a)  \hspace{70mm}  (b)
\end{center}
\caption{ Polarization map and p-I scatter diagram 
 for the original MRN distribution (i.e. $a_{max}=0.25$). 
 We use a logatropic sphere for density, which has a $\rho \propto r^{-1}$
 envelop.
 (a) From the center to the boundary, 
     contours represent 90\%, 50\%, and 10\% of the maximum
     intensity.
 (b) The scatter diagram follows $p \propto I^{1}$ near the center.
 \label{fig:5}
}
\end{figure*}

\begin{figure*}[h!t]
\begin{center}
  \includegraphics[width=0.3\textwidth]{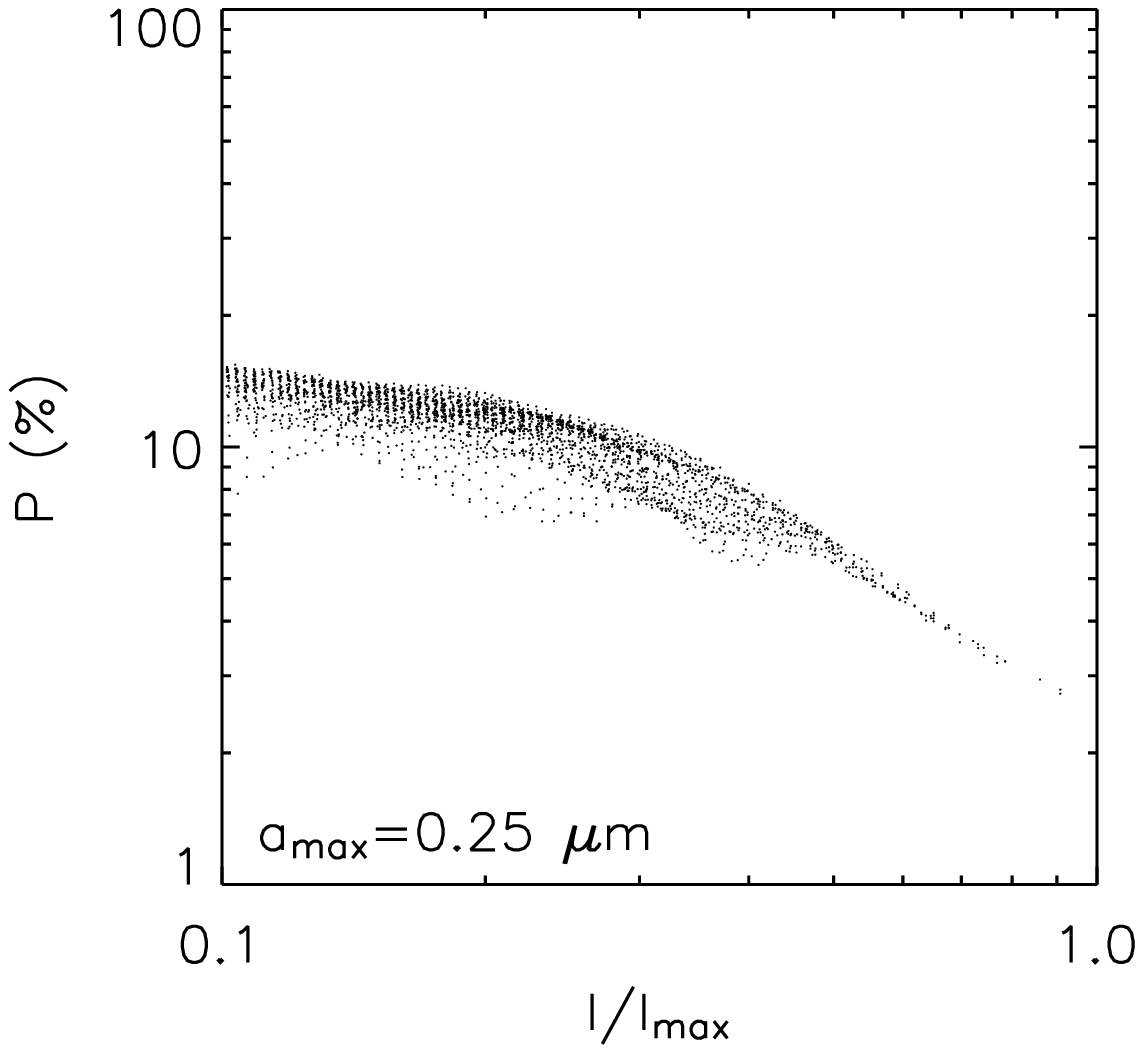} \hspace{3mm}
 \includegraphics[width=0.3\textwidth]{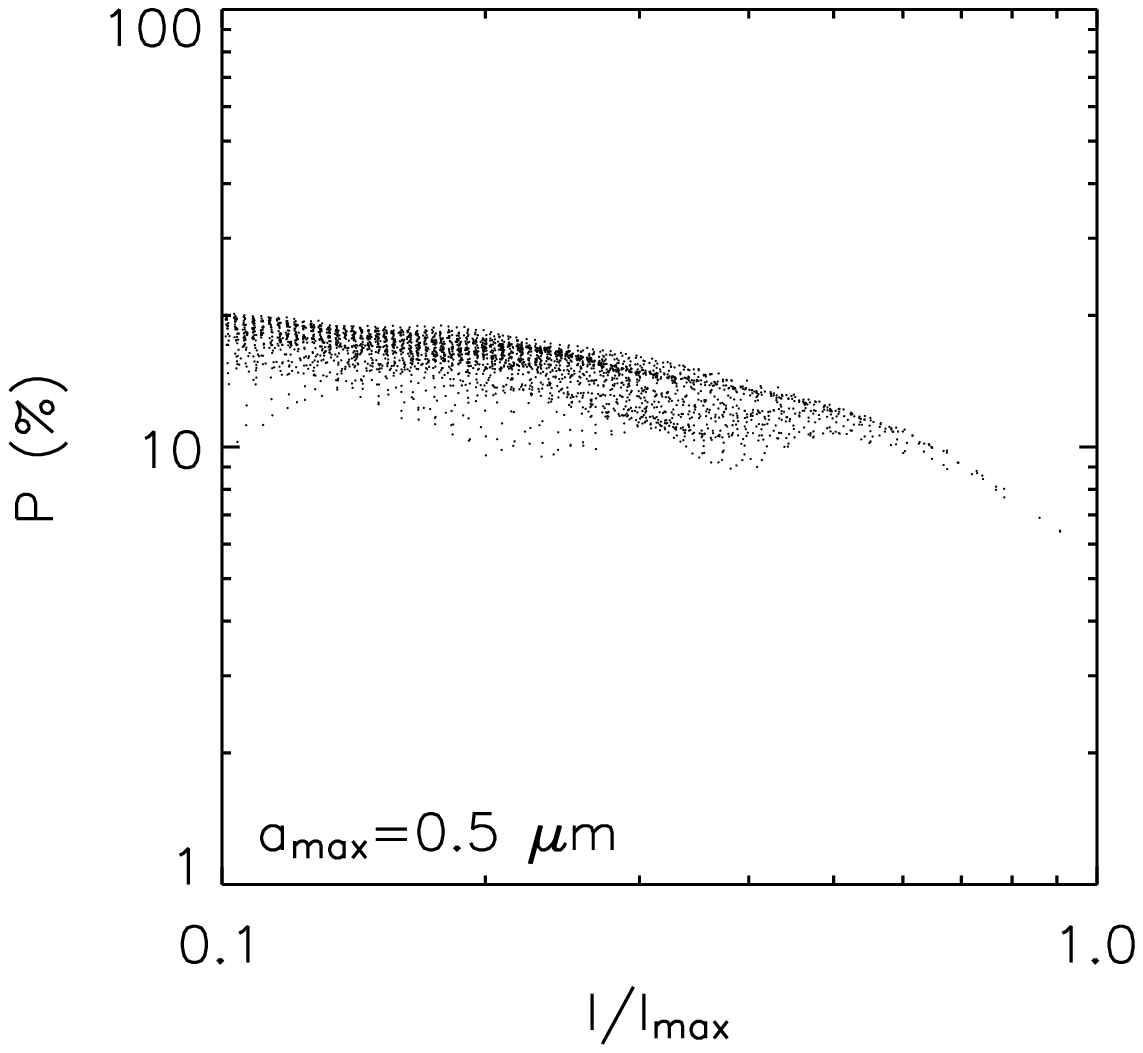}   \hspace{3mm}
 \includegraphics[width=0.3\textwidth]{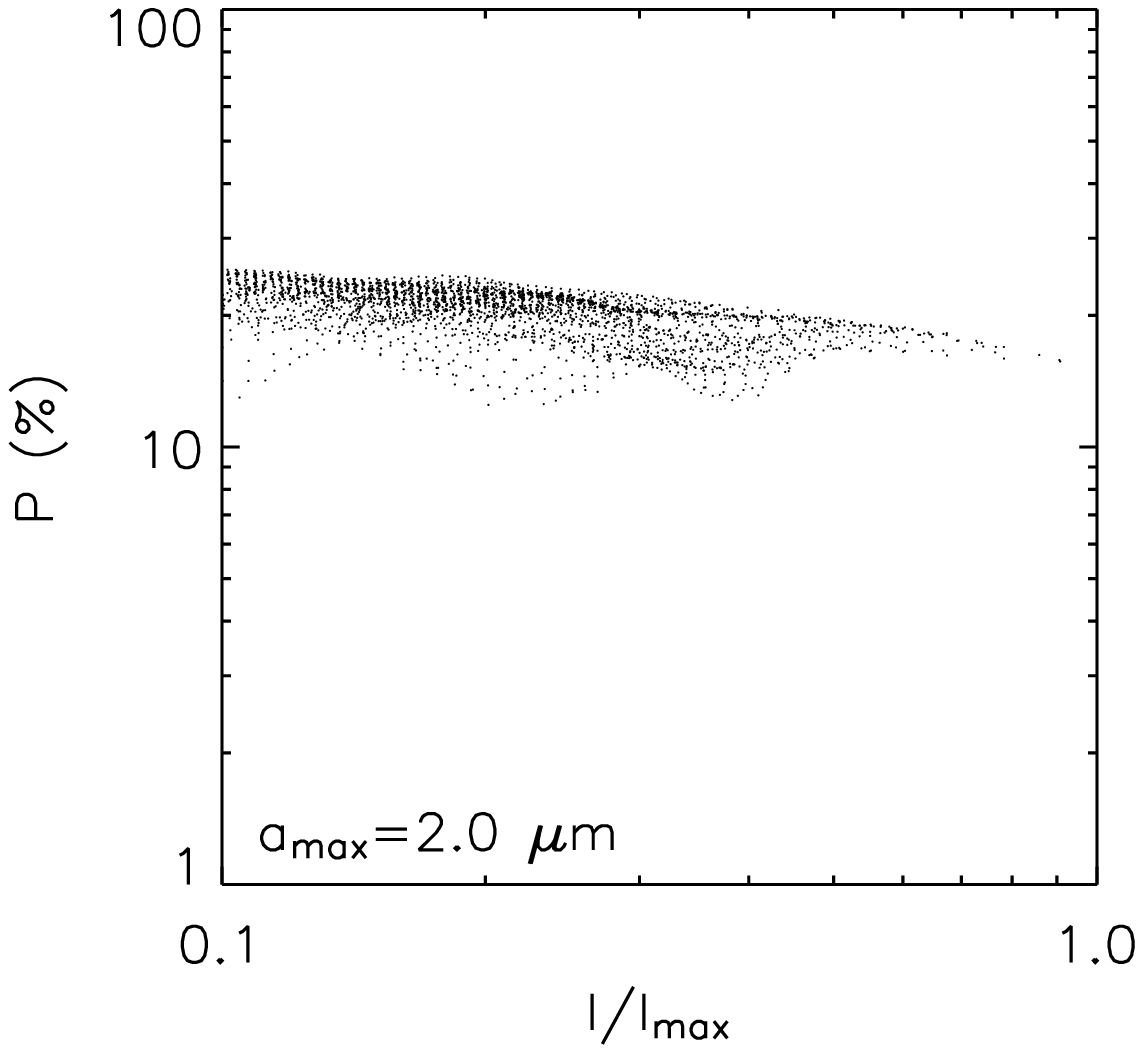} \\
 (a)  \hspace{40mm}  (b)  \hspace{40mm}  (c)
\end{center}
\caption{ The slope gets flatter as the the upper cutoff, $a_{max}$,
          increases. This is because larger cutoff means more aligned
          grains near the cloud center and, therefore,
          higher polarization intensity.
   (a) $a_{max}=0.25\mu$m.   (b) $a_{max}=0.5\mu$m. 
   (c) $a_{max}=2.0\mu$m.
 \label{fig:6}
}
\end{figure*}

\begin{figure*}[h!t]
\begin{center}
\includegraphics[width=.48\textwidth]{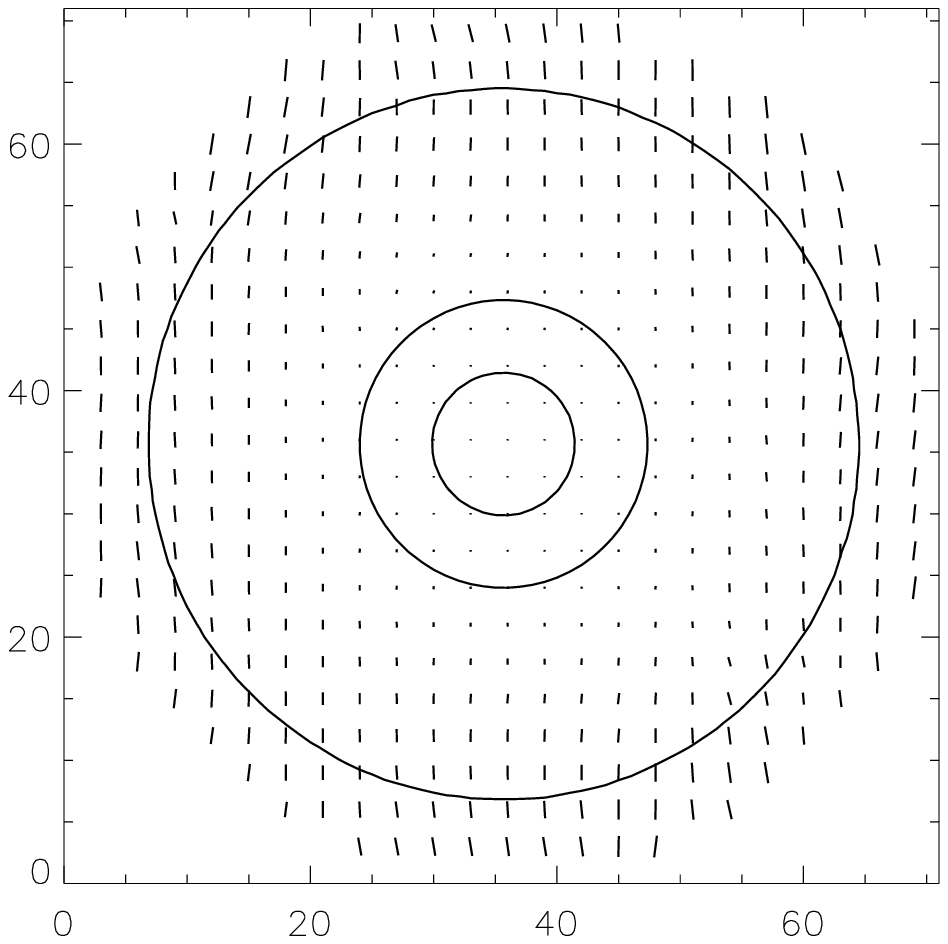}
\includegraphics[width=.48\textwidth]{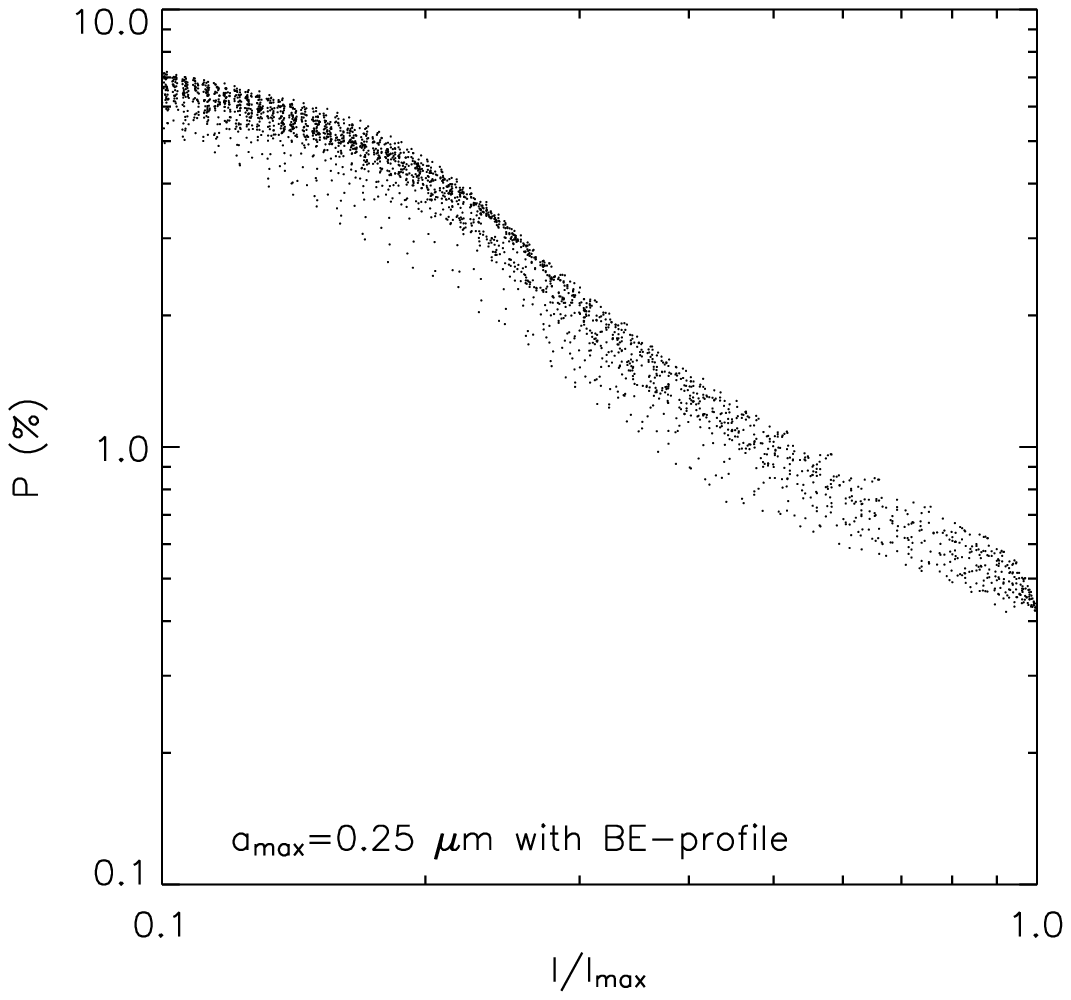}
 (a)  \hspace{70mm}  (b)
\end{center}
\caption{Polarization map and p-I scatter diagram 
 for the original MRN distribution (i.e. $a_{max}=0.25$). 
 We use the isothermal Bonner-Ebert sphere for density, 
 which has a $\rho \propto r^{-2}$ envelop.
  (a) Polarization map.   Contours represent 90\%, 50\%, and 10\% of 
     the maximum intensity.
  (b) p-I relation.
 \label{fig:7}
}
\end{figure*}

\subsection{Simulated map and p-I relation}

In Fig. \ref{fig:5}(a), we plot the polarization map for the original
MRN distribution with $a_{max}=0.25 \mu$m.
We clearly observe the depolarization effect, namely,
the degree of polarization decreases toward the cloud center.
On the other hand, the intensity of dust IR emission is
strongest at the center.
Three contours represent 90\%, 50\%, and 10\% of the maximum
     intensity.
In Figure \ref{fig:5}(b), we plot the degree of polarization ($p$) vs.
intensity ($I$).

Many observations show that the degree of polarization ($p$) decreases
as intensity ($I$) increases.
The relation is usually fitted by a power law.
However, the exact power law index varies.
Matthews \& Wilson (2000) reported $p \propto I^{-0.7}$ for
the OMC-3 region of the Orion A.
On the other hand, Matthews \& Wilson (2002) obtained
$p \propto I^{-0.8}$ for
the dense cores in Barnard 1.
For other dense cores, Henning et al. (2001)
reported $p \propto I^{-0.6}$, Lai et al. (2002) $p \propto I^{-0.8}$,
and Crutcher et al. (2004)  $p \propto I^{-1.2}$.
See Figure 1 of Goncalves, Galli, \& Walmsley (2004) for
illustration. 

We claim that the power law index is sensitive to 
the value of $a_{max}$.
In Figure \ref{fig:6}, we show the change of slopes as a function of 
the upper cutoff $a_{max}$.
The flattening of the slope can be
understood as follows.
When $a_{max}$ gets larger, large grains become relatively more abundant.
Since larger grains are aligned even near the cloud center,
we expect higher degree of polarization near the center.
Therefore, the depolarization effect becomes less pronounced and
the slope gets flatter. We leave for further studies to establish
whether or not the actual slope of the curve can be used to constrain
the grain size distribution.

For the sake of completeness, we also calculate the polarization map
and p-I diagram for the isothermal Bonner-Ebert sphere.
We take the same central density and other parameters 
as in the logatropic sphere.
In Figure \ref{fig:7}(a), we observe more pronounced depolarization effect,
which is because of steeper density gradient.
The p-I scatter diagram in Figure \ref{fig:7}(b) reflects this.

\section{Discussion}

Our calculations show a substantial increase of radiative torque
efficiency as the grain size grows. For the power-law distribution
of grain sizes we have shown that
the p-I relation is sensitive to $a_{max}$ and the density profile
of clouds. Power-law distribution as an approximation for
an actual grain size distirbution was used only for illustrative
purpose. For instance, in Weingartner \& Draine (2001) the 
grain distribution is approximated by a power law up to $a_{max}$
and a exponential tail of grains larger than $a_{max}$. 
It is clear from our calculations that if grains with the size of 
$a_{max}$ are aligned, the grains within the exponential tail
should also be aligned. In fact, the answer to a very important
question whether far infrared polarization reflects the structure of magnetic
field at high optical depths does not depends on the details of the
assumed distribution of grains. It is only essential that a substantial
percentage of dust mass be in sufficiently large grains.  If we want
to predict a polarization spectrum (see Hildebrand et al. 2000) then the
size distribution of grains would matter. For instance, grains of different
sizes can have different temperatures and contribute to either polarized
or not polarized parts of radiation. For  clouds with
embedded stars, the polarization spectrum would depend on the distribution
of stars as well. 

In practical terms our major result is that large grains must be aligned
even at high optical depths. This make sub-millimeter polarimetry
(see Novak et al. 2003)
a useful tool for studies of magnetic fields through the entire process
of star formation. An earlier understanding reflected in, for instance,
Lazarian, Goodman \& Myers (1997) was that embedded stars are essential
for revealing structure of magnetic fields at large optical depth.
This meant that polarimetry might not be able to reveal the role of
magnetic fields at the initial stages of star formation. It worth
mentioning, that observational evidence
 that grains are  aligned in the conditions
when the radiation field is substantially reduced have been recently
claimed to be a major challenge to grain alignment theory (see 
Goncalves, Galli \& Walmsley 2005).

How can we explain that the optical and 
near infrared polarimetry does not detect an appreciable
polarization sygnal originating at high optical depth? We believe
that this stems from the fact that the optical and near infrared  extinction
is biased towards small grains which are not aligned. Qualitatively 
the nature of the bias
can be understood if one recalls that for $\lambda < 2\pi a_c$ the
efficiency of the grains in producing polarized sygnal drops. At the
same time the grains with $a>a_c$ continue to extinct light. Therefore
if a substantial number of grains are larger than $a_c$ the
dichroic properties of the medium in terms of the tranmited light
are affected only by grains from $a_{aligned}$ given by eq.(\ref{eq:ali})
to $\sim a_c$. 
At the same time for grain emission at
$\lambda\gg 2\pi a$ the degree of polarization is determined by the
grains from $a_{aligned}$ to $a_{max}$.  

To illustrate the situation when both near infrared and far infrared
polarimetry would show similar results 
consider a case of the grain alignment at the interface of
the diffuse-dense cloud described in Whittet et al. (2001). For the
range of near-infrared measurements from $0.35~\mu$m to $2.2~\mu$m
the the optical cross section of grains less than $0.25~\mu$m is
still proportional to $a^3$.   
For the case
of the Taurus Dark Clouds Whittet et al. (2001) showed
 that for low optical extinctions,
i.e. $0<A_v<3$ the ratio of the total to selective extinction stays similar
to the value of it in the diffuse gas, i.e. $R_v\approx 3$, while the 
wavelength of maximal polarization $\lambda_{max}$ that enters Serkowski
law (i.e. 
$p_{\lambda}/p_{max}=exp\left[-K ln^2(\lambda_{max}/\lambda)\right]$),
increases. Whittet et al. (2001) interpreted this as the result of the
of size-dependent variations in grain alignment. Lazarian (2003) explained
this as the consequence of the radiative torques which fail to align
small grains at higher optical depths. Our results here support 
this conclusion.
Indeed, if we adopt the grain-size distribution with the original cut-off 
corresponding to the diffuse medium, i.e. $a_{max}=0.25$~$\mu$m,
using our Figure~3 and eq.(\ref{eq:R}) we get Raylegh reduction
factor (or effective alignment) around ten percent at $A_v=3$,
which is a substantial reduction from the value $\approx 1$ for 
$A_v$ of 1. 

At high optical depth the grain-size distributions are rather uncertain. 
To illustrate the importance of grain size growth\footnote{It is important
to realize that the increase of the upper cutoff for the grain size
happens partially due to coagulation of smaller grains. Therefore 
this could be achieved without mantle growth. Naturally, the models
with larger grains, i.e. WD01 do not violate the dust-to-gas ratio.}  for
alignment,
let us use the distribution in WD01 corresponding
to $R_v=5.5$ for $A_v$ of 10 and $n_H=10^4$~cm$^{-3}$. According to 
Fig.~3 only grains with $a>0.6$~$\mu$m are aligned. According to WD01
the favored distribution
of silicate grains is cut-off at a smaller grain size. 
Therefore they are not aligned. The carbonaceous grains have a distribution
with a cut-off
at $\sim$ 1~$\mu$m. As the result we expect $R^{carb}\sim 0.4$, which is
larger that the value of effective alignment for $A_v$ of 3 in the
previous  example\footnote{We expect to have polarization of the level 
of $\approx 3$ percent for this case in emission.}.  
If we use another model of WD01 corresponding to $R_v=5.5$
that has MRN-type distribution of
carbonaceous grains up to size $a\sim 10$~$\mu$m, then the
$R^{carb}$ gets close to $0.8$!
With all these uncertainties we clearly see that far-infrared
polarimetry can get insight into the magnetic  field topology at large
optical depths. In fact, we believe that
far-infrared polarimetry allows consistency checks for
the models of grain-size distributions.

All these results are  valid for clouds without embedded massive stars.
The radiation field is being enhanced in the clouds and therefore we
expect more aligned grains. If grain size distribution stays the same
as in dark clouds, we expect to have high degrees of far-infrared polarization
but still relatively little polarization in terms of optical and near-infrared
polarimetry. The details of this picture  can be tested using polarization
spectrum technique in Hildebrand et al (2000).

We would claim that establishing why some grains are not aligned are as 
important as determining why other grains are aligned. These are two
inseparable parts of the grain alignment problem that must be solved to
make aligned grains a reliable technique for magnetic field study.
On the basis of the present work we believe that we can account for
the polarization arising from dust in dark clouds. Thermal flipping
in the presence of the nuclear spin relaxation described in 
Lazarian \& Draine (1999b) accounts why small grains are not aligned by
Purcell's torques. Therefore we believe that we have a qualitative
agreement between the theory and observations (cf. Goncalves
et al. 2004). We provide a qualitative comparison of the theory and observations in another paper, where we do radiative transfer in a model of a fractal
molecular cloud.

Can the alignment be higher than we predict? Yes, we dealt only with
radiative torques. In fact, calculations in Lazarian \& Draine (1999b)
show that for sufficiently large grains thermal flipping is not
important. As the result such grains are not thermally trapped
and can rotate fast in accordance with Purcell's original predictions.
Naturally, these grains will be aligned paramagnetically. The
requirement for Purcell's torques to work in dark clouds is for
a residual concentration (a fraction of a percent) of atomic hydrogen
to be present or for the grains to have temperatures different from
gas. In addition, MHD turbulence can move grains mostly perpendicular
to magnetic field lines and align them (Yan \& Lazarian 2003). All
these mechanisms are likely to act in unison increasing the alignment
of grains with longer axes perpendicular to magnetic field lines.
 
Our calculations have been motivated by grain alignment in molecular
clouds. Large grains are known to be present in accretion disks
around stars, e.g. protoplanetary disks.
Our work is suggestive that such grains should be aligned and therefore
reflect the structure of magnetic field in disks. As magnetic fields
are believed to play an important role in accretion, the importance
of this is difficult to overestimate. 

The limitation of our calculation is that we used
a magnetic field from a homogeneous MHD simulation without
self-gravity.
In reality, the magnetic field near dense clouds can be very different from
the one we used here.
Recent calculation by Goncalves et al. (2004) shows that
hour glass type magnetic field combined with a torus-like density
profile can cause a depolarized emission from the
cloud center. The reduction factor is around 2.
Therefore, we expect further reduction of polarization
when we use a more realistic magnetic field.

It is also worth mentioning that the radiation fields given in
Mathis et al. (1983) is based on the assumption that
the cloud is spherical and uniform.
Real molecular
clouds are likely to be inhomogeneous and, possibly, hierarchically
clumpy.
As a result, the radiation has
more chances to penetrate deep within molecular clouds
(see Mathis, Whitney, \& Wood, 2002) to allow grains
to be aligned at much higher $A_v$.
Elsewhere we have obtain the radiation field
inside inhomogeneous clouds using direct numerical technique 
similar to the one in Bethell et al. (2004)
and intend to improve our present work by combining our results here and
a more realistic cloud model with realistic radiation field.

The ability to trace magnetic fields inside molecular clouds is difficult
to overestimate. Using Chandrasekhar \& Fermi (1953) technique one can
infer magnetic fields strength both in the cloud and cloud envelope to
test whether star formation takes place in sub or supercritical 
regimes (see Crutcher 2004). The magnetic connection between clouds and
cloud cores is also essential for understanding the processes star
formation. Does magnetic reconnection plays important role for removing
the magnetic flux from molecular clouds? This can be answered 
by studies of magnetic field topology.
In fact, the study of magnetic topology inside molecular clouds could
test different models of magnetic reconnection, e.g. those discussed
in Shay et al. (1998) and Lazarian, Vishniac \& Cho (2004). In addition,
polarimetry
studies of magnetic field structure can bring important insight into the
structure of MHD turbulence insight molecular clouds (see review
by Lazarian \& Yan 2005 and references therein).

Our finding confirm that the present day understanding of grain alignment
can account for all the observational data currently available. This makes
us believe that polarization arising from aligned grains has become a tool
based on solid theoretical foundations. The latter is important not only for
molecular cloud studies but for many other studies, e.g. those of comet,
circumstellar polarimetry (see Lazarian 2003 and references therein)
as well as for predicting and enterpolating to other wavelength the
polarized foreground contribution from dust (see review by
Lazarian \& Finkbeiner 2004 and references therein).

\section{Summary}

We have studied the efficiency of grain alignment by radiative torque 
in optically thick clouds. We have estabilished that the efficiency
 of radiative
torques is a steep function of the grain size. As the result,
even deep inside giant molecular clouds ($A_V \lesssim 10$), 
large grains can be aligned by radiative torque. This means that
far-infrared/submillimeter polarimetry can reliably reflect the
structure of magnetic field deep inside molecular clouds.
Our results show that the grain size is important 
for determining the relation between the degree of polarization and 
intensity from molecular cloud dust.

\begin{acknowledgments}  
This work is supported by NSF grant AST-0243156 and the NSF Center for
Magnetic Self-Organization in Laboratory and Astrophysical Plasmas.
This work utilized CITA supercomputing facilities during its early stages.
We thank Bruce Draine, 
Dick Crutcher, Roger Hildebrand, John Mathis,
and Giles Novak for useful discussions.
\end{acknowledgments}

\end{document}